\documentclass[conference]{IEEEtran}
\IEEEoverridecommandlockouts
\usepackage{cite}
\usepackage{amsmath,amssymb,amsfonts}
\usepackage{algorithmic}
\usepackage{graphicx}
\usepackage{textcomp}
\usepackage{xcolor}
\def\BibTeX{{\rm B\kern-.05em{\sc i\kern-.025em b}\kern-.08em
    T\kern-.1667em\lower.7ex\hbox{E}\kern-.125emX}}
\begin{document}

\title{AiEDA: Agentic AI Design Framework for Digital ASIC System Design}

\author{\IEEEauthorblockN{Aditya Patra}
\IEEEauthorblockA{
\textit{California High School}\\
San Ramon, USA \\
adityabpatra@gmail.com}
\and
\IEEEauthorblockN{Saroj Rout}
\IEEEauthorblockA{\textit{Dept. of Electronics Engineering} \\
\textit{Silicon University, Odisha}\\
Bhubaneswar, India \\
ORCID: 0000-0002-5191-8191 }
\and
\IEEEauthorblockN{Arun Ravindran}
\IEEEauthorblockA{\textit{Dept. of Electrical and Computer Engineering} \\
\textit{University of North Carolina at Charlotte}\\
Charlotte, USA \\
arun.ravindran@charlotte.edu }
}

\maketitle

\begin{abstract}
\label{sec:abstract}
	The paper addresses advancements in Generative Artificial Intelligence (GenAI) and digital chip design, highlighting the integration of Large Language Models (LLMs) in automating hardware description and design. LLMs, known for generating human-like content, are now being explored for creating hardware description languages (HDLs) like Verilog from natural language inputs. This approach aims to enhance productivity and reduce costs in VLSI system design. The study introduces ``AiEDA," a proposed agentic design flow framework for digital ASIC systems, leveraging autonomous AI agents to manage complex design tasks. AiEDA is designed to streamline the transition from conceptual design to GDSII layout using an open-source toolchain. The framework is demonstrated through the design of an ultra-low-power digital ASIC for KeyWord Spotting (KWS). The use of agentic AI workflows promises to improve design efficiency by automating the integration of multiple design tools, thereby accelerating the development process and addressing the complexities of hardware design.
\end{abstract}


\section{Introduction}
\label{sec:introduction}

The development of Generative AI (GenAI) Large Language Models (LLMs), coupled with the easy API based accessibility of powerful models with hundreds of billions of parameters,  has brought significant advancements in artificial intelligence, allowing systems to generate human-like content across various mediums, including text, images, and code \cite{intro2LLM}. Meanwhile, digital chip design is becoming increasingly complex, requiring management of millions to billions of transistors while optimizing performance, power consumption, and area. Furthermore, it demands precise coordination of design factors such as timing, signal integrity, and manufacturability, all within stringent time-to-market constraints. Recent research has investigated the application of LLMs in digital design, particularly in generating hardware description languages (HDLs) such as Verilog from natural language design descriptions. The use of LLMs in VLSI system design is part of a larger effort by the research community to improve designer productivity, and costs needed to realize complex System-on-Chips (SoCs) \cite{ajayi2019openroad}.

\begin{figure}[htbp]
	\includegraphics[width=0.45\textwidth]{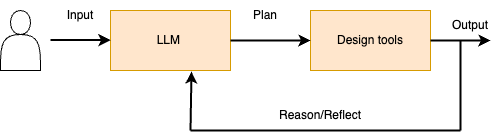}
	\caption{Agentic AI workflow involves an iterative process involving one or more LLMs for reasoning, and planning, and one or more external tools to execute actions. The input could be a design specification in natural language, or a behavioral description in an HDL. The output could be HDL, netlist, or a GDS layout.
}
	\label{fig:agentic_overview}
\end{figure}

In the last year or so, agentic AI workflows have emerged, where autonomous AI agents perform specific tasks within defined parameters \cite{promptengineering2024}. Agentic AI systems are defined by their capacity to take actions that consistently work toward achieving goals over time, even when their behavior is not pre-programmed in advance. Agents utilize an LLM to reason and determine the sequence of actions to take. Figure \ref{fig:agentic_overview} shows the high-level outline of an agentic workflow.  These workflows are increasingly used in software design to automate tasks such as code generation, debugging, and testing, thus improving development speed and minimizing human error. While agentic AI workflows have proven effective in software code generation \cite{alphacodium}, applying these techniques to hardware design presents additional complexities. This is due to the diverse range of tools required to meet functional and timing correctness, in addition to physical constraints in hardware systems.

In this paper, we propose AiEDA, an agentic design flow framework in the design of a digital ASIC system from concept to GDSII using an open source tool flow. Our premise is that the use of agentic AI workflows in digital design would greatly increase designer productivity, enabling rapid implementation of design ideas that integrate a number of different design tools, to generate a full system design is ready to be fabricated. We demonstrate the use of the proposed framework, via the design of a ultra low power digital ASIC for KeyWord Spotting(KWS) architecture. 

The paper is organized as follows. In Section \ref{sec:related_work}, we briefly review the literature on the use of generative AI in hardware design. 
In Section \ref{sec:agentic}, we describe AiEDA, our proposed agentic AI design framework. In Sections \ref{sec:system_design} and Sections \ref{sec:preliminary_evaluation} we present the KWS architecture, and our initial results in using AiEDA to design it. In Section \ref{sec:discussion}, we discuss the broader applicability of agentic flow in digital system design, the open research questions that need to be addressed, and our current research efforts in this direction. Section \ref{sec:conclusions} concludes the paper.

\section{Related Work}
\label{sec:related_work}

\begin{figure*}[htbp]
    \centering
    \includegraphics[width=0.6\textwidth]{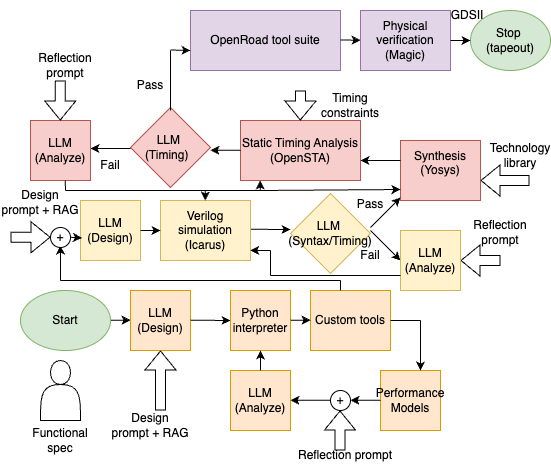} 
    \caption{Proposed agentic AI design framework for digital ASIC design. The design flow is broadly divided into four parts: 1) Architecture design (shaded orange), 2) RTL design (shaded yellow), 3) Netlist synthesis (shaded red), and 4) Physical design (shaded purple). At each stage, the design is driven by a combination of LLM and appropriate EDA tools operating in a feedback loop.  }
    \label{fig:agentic_designflow}
\end{figure*}
We review previous reported work on the use of Generative AI in chip design. DAVE \cite{dave} and Verigen \cite{verigen} were among the first works to exploit the use of LLMs in HDL generation. Their focus is on improving LLM performance by fine-tuning the open source LLM models for Verilog code generation. 

The ChipNemo project \cite{chipnemo} from Nvidia does not directly target the use of LLMs in the design of digital systems. Instead, domain-specific LLMs are designed for supporting ancillary tasks such as a chatbot that serves as an engineering assistant, EDA script generation, and bug summarization and analysis. Their efforts are thus complementary to our proposed work. 

Recognizing the need to augment and fine-tune LLMs for hardware design, the MG-Verilog \cite{mg-verilog} project proposes an open source dataset consisting of over 11,000 Verilog code samples and their corresponding natural language descriptions. MG-Verilog dataset is complementary to our effort, and could be used to make the underlying foundational models more accurate through techniques such as Retrieval Augmented Generation (RAG), and fine tuning.  

In the GPT4AIChip project \cite{gpt4aigchip}, a feedback design loop is proposed consisting of prompting the LLM with human-crafted design examples (few-shot learning) and an evolutionary algorithm-based design space exploration algorithm to explore the designs generated by the LLM.  The LLM generated code for a matrix multiplier (GEMM) is synthesized and evaluated on an FPGA using the Xilinx Vivado HLS tools. The design flow proposed by GPT4AIChip can be considered agentic, but its design exploration space is constrained. In contrast to our approach, it does not address the more complex ASIC design flow, including synthesis and timing analysis.

The AutoChip \cite{autochip} project proposes an automated approach that uses large language models (LLMs) to generate HDL. AutoChip combines conversational LLMs such as OpenAI GPT-4 with feedback from Verilog compilers and simulations to iteratively improve Verilog modules. Starting with an initial module generated from a design prompt, it refines the design based on errors and simulation messages. AutoChip's effectiveness is evaluated using design prompts and test benches from HDLBits, a collection of small circuit design exercises. Although AutoChip uses an agent-based design flow, unlike our system's design focus, it is limited to small circuits and lacks a comprehensive end-to-end design flow.

\section{Proposed Agentic AI Design Framework}
\label{sec:agentic}

In this Section, we describe AiEDA, our proposed agentic AI based design framework for digital ASIC system design. Figure \ref{fig:agentic_designflow} shows the proposed design  framework. The design flow is broadly divided into four interrelated stages: 1) Architecture design (shaded orange), 2) RTL design (shaded yellow), 3) Netlist synthesis (shaded red), and 4) Physical design (shaded purple). At each stage, the design is driven by a combination of LLM and appropriate EDA tools operating in a feedback loop. Inputs to the LLMs include design prompts, which instructs the LLM to follow a particular set of actions, or reflection prompts which instructs the LLM to analyze the results of an action. Additionally, the capabilities of LLMs are enhanced with Retrieval Augmented Generation (RAG) techniques, where retrieval of appropriate knowledge (for example, Verilog code dataset such as MG-Verilog) provides a richer context to the LLM in generating a more accurate response. Furthermore, custom LLMs could be employed that have been fine-tuned with domain-specific datasets.

During the architecture design phase, the developer starts with a high-level system specification. They use a large language model (LLM) to decompose the system into individual components and generate a Python model representing these components. This Python script interacts with APIs for custom tools (e.g., MATLAB toolboxes via the MATLAB Engine API) and performance models. The results are then analyzed by another LLM to inform adjustments to the overall system design. This iterative design loop is repeated to evaluate trade-offs between design metrics, such as area and accuracy. The developer can intervene at any point by refining the LLM prompts to guide the design process according to their expertise.

In the RTL design phase, the architectural design in Python, design prompts from the designer, and Verilog code data generated using Retrieval Augmented Generation (RAG) are provided to an LLM, which generates Verilog RTL for the design along with the necessary test benches. Functional verification is performed using the open-source Icarus simulator, and the output is analyzed by the LLM. The LLM identifies failures and generates the necessary modifications to the Verilog code. This feedback loop is repeated until functional verification is successful. As in the architectural design phase, the designer can intervene at any point to modify the prompts, Verilog RTL, or test benches to guide the design toward a desired outcome.

In the Netlist synthesis phase, the Verilog RTL code is synthesized into a gate-level netlist using the open-source Yosys synthesizer, based on standard logic gates from the target process technology library. Timing analysis is then performed using the open-source OpenSTA static timing analyzer. Any timing violations are analyzed by an LLM, and a second LLM generates corrective actions to optimize the timing paths. This is followed by re-running the synthesis with adjusted constraints or updated RTL. 

In the Physical design phase, the design's physical layout is created using tools from the open-source OpenROAD tool suite for backend tasks such as placement, clock tree synthesis, routing, and optimization \cite{ajayi2019openroad, ajayi2019toward}. OpenROAD utilizes multiple feedback mechanisms, including Design Rule Check (DRC) feedback, timing analysis feedback, power analysis feedback, and area feedback. Integrating agentic AI into these feedback loops is a potential area for future exploration. The final step uses the open-source Magic tool to generate GDSII files, which are then used for chip fabrication.

\section{Case study - Design of Digital Keyword Spotter}
\label{sec:system_design}

\begin{figure}[htbp]	
    \includegraphics[width=0.45\textwidth]{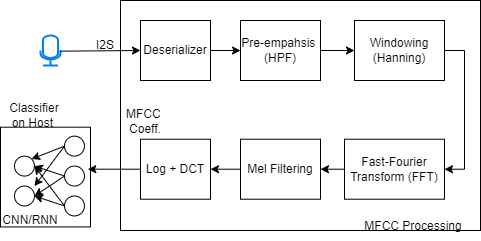}
    \caption{Keyword Spotter (KWS) architecture.}
    \label{fig:KWS_Arch}
\end{figure}

In this Section, we describe a case study of the application of our proposed design framework to the design of a digital keyword spotter system. 

Acoustic audio analysis is widely applied in critical areas such as heart monitoring, keyword spotting (KWS), and the preventive maintenance of heavy machinery \cite{chong20220}. Among these applications, KWS is essential for voice-activated assistants like Google Assistant and Amazon Alexa.

Figure~\ref{fig:KWS_Arch} depicts a widely recognized digital KWS architecture \cite{chong20220}. In this design, Mel frequency cepstral coefficients (MFCC) are used to extract spectral features from the input audio signal. 

A digital microphone captures live audio data via an I²S serial interface, which is then converted into parallel bytes. A pre-emphasis filter (high-pass filter or HPF) removes the low-frequency components of the signal. The pre-emphasis filter is typically expressed by the difference equation:
\begin{equation}
    y[n] = x[n] - \alpha~x[n-1] \label{eq:hpf}
\end{equation}
where $\alpha$ typically ranges from 0.9 to 1 \cite{han2006efficient}.

Next, a window function (e.g., Hamming or Hanning) is applied to the data to minimize spectral leakage during the FFT process. The fast Fourier transform (FFT) is then used to extract the signal’s frequency components. However, since the linear frequency scale provided by FFT does not correspond to how humans perceive sound, the linear spectrum is transformed into a Mel scale using the following equation \cite{han2006efficient}:
\begin{equation}
    Mel(f) = 2595\cdot \log_{10}\left({1 + \frac{f}{700}}\right) \label{eq:mel-log}
\end{equation}

After applying the Mel-scale transformation, the logarithm of the Mel frequency power is computed. A discrete cosine transform (DCT) is subsequently used to generate MFCC coefficients. These coefficients are then fed into a classifier, such as a convolutional neural network (CNN), recurrent neural network (RNN), or long short-term memory (LSTM) network, to recognize the keyword \cite{mahmood2021speech}. Since the classifier is hardware-intensive and computationally demanding, it is most efficiently executed on the host processor.
\section{Preliminary Evaluation}
\label{sec:preliminary_evaluation}
In this section, we present a preliminary evaluation of the Keyword Spotting (KWS) architecture using our proposed AiEDA design flow. It is important to note that this work is still ongoing. AiEDA is implemented in Python, with the agentic flow built using the LangGraph framework from LangChain \cite{langgraph}. The design tools referenced are those described in Section \ref{sec:agentic}. The Large Language Model (LLM) utilized in this evaluation is OpenAI's GPT-4o model. As the design tools are still in active development, we plan to release the project as open-source once the code has reached a stable and mature state.

\subsection{AiEDA - Design specification to GDS}

Our initial goal was to validate the AiEDA design flow from RTL to GDS using a relatively simple component. The RTL, synthesis, and netlist stages of AiEDA (refer to Figure Section \ref{sec:agentic}) were tested by designing a 6-bit, 32-depth FIFO. The design process began with a prompt specifying the FIFO's requirements and concluded with the generation of the GDSII layout. The Sky130 standard cell library and PDK from SkyWater Technologies were utilized for this implementation.

\subsection{AiEDA - Architecture design}
We note that despite the popularity of the KWS architecture, hardware implementation can differ greatly depending on the application, which involves various trade-offs between power, area, and accuracy.
For this work, a \textit{smart microphone } application was examined in which the KWS is intended to be embedded directly into the microphone. This configuration allows it to recognize one or more keywords with moderate accuracy, consuming minimal power and occupying a small footprint. Each component is evaluated to determine how its design can be optimized for the specific application.

Digital microphones are generally built to record audio frequencies up to 22~kHz with a sample precision ranging from 12-16 bits. The initial design prompt captured the requirements, and instructed the LLM to explore the bandwidth and precision requirements. The open-source audio tool \textit{Audacity} was incorporated into the design flow using the \textit{ mod-script-pipe} plugin. The reflection prompt directed the LLM to optimize the bandwidth so that the total power of the input signal retains at least 90\% of its original power. Additionally, it instructed the LLM to optimize the bit-width precision ensuring that the tonal component (derived from the FFT operation) within the bandwidth remains largely unaffected.
 After multiple iterations, a 4 kHz bandwidth and 7-bit fixed-point precision were found sufficient for the operation. The Nyquist sampling frequency, which is twice the bandwidth, will decrease from 44 kHz to 8 kHz, leading to an approximately 5x reduction in power consumption. Additionally, using 7-bit precision will yield around a 2x reduction in both area and power consumption.

Next, we designed the individual architectural components shown in Figure \ref{fig:KWS_Arch}. An iteration of the design similar to the overall system was performed for the HPF. The LLM was instructed (Eq.~\ref{eq:hpf}) to select an $\alpha$ that can be implemented using a simple shift-and-add operation, eliminating the need for a hardware multiplier. Following reflection by the LLM, the final design chose an $\alpha = 31/32 = 0.969$ that resulted in an insignificant DC component (first 1-2 bins of the FFT) without a significant loss of audio spectrum. 

A Hanning window was selected for the windowing function, and the LLM was instructed to design approximate coefficients exclusively using single-shift operations such that the spectral leakage is limited to 10\%. The output spectrum was analyzed using a Python based script. Following reflection, the LLM precomputed the coefficients in fixed-point form, generating values that can be efficiently approximated by bit-shifts (i.e., powers of 2).

FFT is the largest hardware component with respect to area and power consumption. The $Radix-2^2$ single delay feedback ($R2^2SDF$) is a design that is efficient in both area and power usage, since it uses the fewest multipliers and adders compared to other FFT hardware implementations \cite{chong20220}. The output spectrogram was analyzed using a Python based script. The LLM was tasked with evaluating FFT sizes ranging from 16-point to 256-point to identify the minimum number of points that would limit the loss of accuracy to within 25\%. Upon analysis of the spectrogram, the LLM determined that a 32-point FFT provides fewer frequency bins while achieving substantial hardware savings.

Similar hardware reduction strategies were achieved by employing rectangular Mel filter bins as opposed to triangular ones, with only a slight reduction in accuracy. The final component, the discrete cosine transform (DCT), is a multiply-accumulate (MACC) function \cite{chong20220}. Similar to the Hanning window design procedure, the coefficient multiplications are transformed into a \textit{shift-and-add} method to reduce the complexity of the hardware without greatly affecting the accuracy.

 The host microcontroller then utilizes these MFCC coefficients with neural network-based classifiers such as CNN or RNN to spot the trained keyword.

\subsection{AiEDA - End-to-end design}
Building on the architectural design of the KWS system, we are now working on developing the full end-to-end system. This process starts with the architectural description in Python and will culminate in the GDSII output. Our goal is to complete this before the conference date.
\section{Discussion}
\label{sec:discussion}
We have introduced an agentic AI design framework aimed at boosting productivity in the development of complex digital ASICs. This framework is still in its early stages, but we plan to present more detailed evaluation results and release an open-source prototype of AiEDA by the conference.

In developing AiEDA, we faced several challenges. One key issue is whether to use a powerful general-purpose LLM like GPT-4o or to customize smaller models for specific tasks. The decision involves balancing customization and deployment costs against potential gains in accuracy and reduced operational expenses.

Another crucial consideration is the role of the designer. While it's possible to argue for an AI tool that enables novice designers to create digital ASICs independently, we believe AI should enhance human creativity rather than replace it. AiEDA is designed to allow designers to intervene and guide the process based on their expertise at any stage (Reinforcement Learining with Human Feedback). 

Future areas of exploration include creating an open-source library of design prompts and integrating optimization tools into the AI framework.
\section{Conclusions}
\label{sec:conclusions}
In this paper, we have introduced AiEDA, an agentic AI design framework for digital ASIC design. AiEDA leverages advanced Generative AI techniques, including agentic workflows that integrate open-source EDA tools, prompt engineering, few-shot learning, self reflection, and retrieval-augmented generation. We have provided initial evaluation results based on the design of a KeyWord Spotter system. As this work continues to develop, we anticipate presenting more comprehensive evaluation findings by the conference.


\bibliographystyle{IEEEtran}
\bibliography{references}

\end{document}